\documentclass[twocolumn,showpacs,preprintnumbers,amsmath,amssymb,nofootinbib]{revtex4}

\usepackage{graphicx}
\usepackage{dcolumn}
\usepackage{bm}

\begin{document}

\preprint{APS/123-QED}

\title{Competitive effects of nuclear deformation and density dependence of
$\Lambda\!N$ interaction \\
in $B_\Lambda$ values of hypernuclei}

\author{M.\ Isaka$^{1}$}
\author{Y.\ Yamamoto$^{2}$}
\author{Th.A.\ Rijken$^{3}$$^{2}$}
\affiliation{
$^{1}$Research Center for Nuclear Physics (RCNP), Osaka University, Ibaraki, Osaka, 567-0047, Japan\\
$^{2}$Nishina Center for Accelerator-Based Science,
Institute for Physical and Chemical
Research (RIKEN), Wako, Saitama, 351-0198, Japan\\
$^{3}$IMAPP, University of Nijmegen, Nijmegen, The Netherlands
}

\date{\today}

\begin{abstract}
Competitive effects of nuclear deformation and density dependence of
$\Lambda\!N$-interaction in $\Lambda$ binding energies $B_\Lambda$ of hypernuclei are studied systematically
on the basis of the baryon-baryon interaction model ESC including many-body effects. 
By using the $\Lambda\!N$ G-matrix interaction derived from ESC, we perform microscopic calculations
of $B_\Lambda$ in $\Lambda$ hypernuclei within the framework of the antisymmetrized molecular 
dynamics under the averaged-density approximation. The calculated values of $B_\Lambda$ reproduce 
experimental data within a few hundred keV in the wide mass regions from 9 to 51. 
It is found that competitive effects of nuclear deformation and density dependence of
$\Lambda\!N$-interaction work decisively for fine tuning of $B_\Lambda$ values.
\end{abstract}

\pacs{Valid PACS appear here}
\maketitle

\section{Introduction}

Basic quantities in hypernuclei are $\Lambda$ binding energies 
$B_\Lambda$, from which a potential depth $U_\Lambda$ 
in nuclear matter can be evaluated.
The early success to reproduce the $U_\Lambda$ value was achieved by
Nijmegen hard-core models~\cite{NDF}, where the most important 
role was played by the $\Lambda N$-$\Sigma N$ coupling term. 
Medium and heavy $\Lambda$ hypernuclei have been 
produced by counter experiments such as $(\pi^+,K^+)$ reactions.
Accurate data of $B_\Lambda$ values in ground and excited states of
hypernuclei have been obtained by $\gamma$-ray observations and
$(e,e'K^+)$ reactions. With the increase of experimental information 
\cite{PPNP57.564(2006)}, precise interaction models have been constructed.
In the Nijmegen group, the soft-core models have been developed with 
continuous efforts so as to reproduce reasonably hypernuclear 
data~\cite{NSC89,NSC97,ESC04,ESC08}.
In the recent versions of the Extended-Soft Core (ESC) models~\cite{ESC04,ESC08},
two-meson and meson-pair exchanges are taken into account explicitly, while 
these effects are implicitly and roughly described by exchanges 
of ``effective bosons" in one-boson exchange (OBE) models.
The latest model ESC08c aims to reproduce consistently almost 
all features of the $S=-1$ and $-2$ systems. The parameter fitting 
has been improved continuously, and the final version has to be submitted soon.
In Ref.~\cite{PRC89.024310(2014)}, they used successfully 
the version of 2012 in the early stage of parameter fitting~\cite{ESC2012}, denoted as ESC08c(2012) \footnote{The fortran code ESC08c2012.f is put on the permanent open-access website, NN-Online facility: {\it http://nn-online.org}.}.
This version is used also in the present work.

Recently, the dependence of $B_\Lambda$ on structures of core nuclei, in particular nuclear deformations, has been discussed in $p$ shell \cite{PRC92.044326(2015)}, and $sd$-$pf$ shell hypernuclei \cite{PRC89.024310(2014),PRC89.044307(2014),PTEP103D02(2015)} theoretically. Generally, values of $B_\Lambda$ are related to nuclear structure in two ways:
One is that an increase of deformation reduces the overlap of the densities between a $\Lambda$ particle and the core nucleus, which makes $B_\Lambda$ smaller. 
Such effects are seen in $sd$-$pf$ shell hypernuclei. In Refs.~\cite{PRC89.024310(2014),PTEP103D02(2015)}, the antisymmetrized molecular dynamics for hypernuclei (HyperAMD) \cite{PRC83.054304(2011),PRC85.034303(2012)} was applied to several $sd$-$pf$ shell hypernuclei such as $^{41}_\Lambda$Ca and $^{46}_\Lambda$Sc. It is found that $B_\Lambda$ values in deformed states are decreased reflecting smaller overlaps. 

The other effect is due to the density dependence of the $\Lambda N$ effective interaction. 
In light hypernuclei and/or dilute states like cluster states, the density overlap between a $\Lambda$ and nucleons is significantly decreased, which can affect the $B_\Lambda$ through the density dependence. 
For example, in Be hypernuclei having a 2$\alpha$ cluster structure with surrounding neutrons, it was discussed that the overlap becomes much smaller in the well-pronounced 2$\alpha$ cluster states \cite{PRC92.044326(2015)}. 
When the $\Lambda N$ effective interaction derived from the $G$-matrix calculation is designed to depend on the nuclear Fermi momentum $k_F$, the smaller overlap makes the relevant value of $k_F$ small, {\it i.e.} less Pauli-blocking, resulting in the increase of $B_\Lambda$. 
Considering this effect, it is expected that appropriate values of $k_F$ in finite systems are 
reduced as overlaps become small with mass numbers, which would affect 
the mass dependence of $B_\Lambda$. 

$\Lambda N$ interactions are related intimately to the recent topic of heavy neutron stars (NS). 
The stiff equation of state (EoS) giving the large NS-mass necessitates the strong 
three-nucleon repulsion in the high-density region, the existence of which has been established
by many works \cite{UIX} in nuclear physics. However, the hyperon mixing in neutron-star matter 
brings about the remarkable softening of the EoS, canceling this repulsive effect. 
A possible way to solve such a problem is to assume that strong repulsions 
exist universally in three-baryon channels. More specifically, it is assumed that 
the $\Lambda NN$ repulsion works in $\Lambda$ hypernuclei as well as the
three-nucleon repulsion.
A $\Lambda NN$ three-body effect, that is generally a hyperonic many-body effect (MBE),  
has to appear as an additional density dependence of the $\Lambda N$ effective interaction.
It is important to study MBE by analyzing the experimental data of $B_\Lambda$ systematically.

The aim of the present work is to reveal how the density dependence of the $\Lambda N$ effective interaction affects the mass dependence of $B_\Lambda$. 
Since the $p$-$sd$-$pf$ shell hypernuclei have various structures in the ground states, they would affect the values of $B_\Lambda$ through the density-dependence of the $\Lambda N$ interaction. To investigate it, we use the HyperAMD combined with $\Lambda N$ $G$-matrix interaction, which successfully describes various structures of hypernuclei without assumptions on specific clustering and deformations \cite{PRC83.054304(2011),PRC85.034303(2012)}. 

This paper is organized as follows. In the next section, the $\Lambda N$ G-matrix interaction is explained as well as treatment of MBE. In Sec. III, we explain how to describe hypernuclei, namely the theoretical framework of HyperAMD. In Sec. IV, we show the calculated values of $B_\Lambda$ including MBE, and 
discuss effects from core structures on $B_\Lambda$ . 
Section V summarizes this paper.

\section{$\Lambda N$ G-matrix interaction}
\label{SecII}

We start from ESC08c(2012),
which was used in the analysis of $\Lambda$ hypernuclei
based on the HyperAMD most successfully \cite{PRC89.024310(2014)}.
One should be careful, however, that the main conclusion in this work
has to be valid qualitatively also for other realistic interaction models
including $\Lambda N$-$\Sigma N$ coupling terms which lead to strong
density dependences of the $\Lambda N$ effective interactions.
Hereafter, ESC08c(2012) is denoted as ESC simply.
As a model including an additional density dependence due to a hyperonic MBE,
we adopt the model given in Ref.~\cite{YFYR14}.
Here, the multi-pomeron exchange repulsion (MPP) is added into ESC
together with the phenomenological three-body attraction (TBA),
where both of them are represented as density-dependent two-body interactions.
Using ESC+MPP+TBA, G-matrix calculations are performed with the continuous choice
for off-shell single particle potentials: Contributions of MPP and TBA are 
renormalized into $\Lambda\!N$ G-matrices.
The MPP part is given as
\begin{eqnarray}
&& V_{MPP}^{(N)}(r;\rho) 
 \nonumber \\
&& 
=g_P^{(N)} g_P^N\frac{\rho^{N-2}}{{\cal M}^{3N-4}}
 \left(\frac{m_P}{\sqrt{2\pi}}\right)^3
 \exp\left(-\frac{1}{2}m_P^2 r^2\right),
\label{eq:2}
\end{eqnarray}
corresponding to triple ($N=3$) and quartic ($N=4$) pomeron exchange.
The values of the two-body pomeron strength $g_P$ and 
the pomeron mass $m_P$ are the same as those in ESC.
A scale mass ${\cal M}$ is taken as the proton mass.
The TBA part is assumed as
\begin{eqnarray}
&& V_{TBA}(r;\rho)
\nonumber \\
&& = V_{0}\, \exp(-(r/2.0)^2)\, \rho\, 
\exp(-3.5 \rho)\, (1+P_r)/2 \ ,
\label{eq:4}
\end{eqnarray}
$P_r$ being a space-exchange operator.
In Refs.~\cite{YFYR13,YFYR14}, these interactions were assumed to be
universal in all baryonic channels. Namely, the parameters
$g_P^{(3)}$, $g_P^{(4)}$ and $V_0$ in hyperonic channels were
taken to be the same as those in nucleon channels,
assuring the stiff EoS of hyperon-mixed neutron-star matter.
There were used three sets with different strengths of 
MPP in Refs.~\cite{YFYR13,YFYR14}. In the case of the set MPa,
for instance, the parameters were taken as
$g_P^{(3)}=2.34$, $g_P^{(4)}=30.0$ and $V_0=-32.8$.
In the present analysis, however, such a choice leads to a too strong density-dependence of
the $\Lambda N$ G-matrix interaction for reproducing the mass dependence of $B_\Lambda$ values:
In the case of ESC08c(2012), the mass dependence of $B_\Lambda$ values are reproduced 
rather well without the additional MBE.
Then, the values of $g_P^{(3)}$ and $g_P^{(4)}$ may be much smaller than 
the above values so that the additional density dependence is not strong.
Here, the parameters are determined so that calculated results of 
$B_\Lambda$ values in the present framework are consistent with the experimental data.
They are taken as $g_P^{(3)}=0.39$, $g_P^{(4)}=0.0$ and $V_0=-5.0$:
MPP (TBA) is far less repulsive (attractive) than those in the above case. 
In this case, the calculated value of $B_\Lambda$ is 13.0 MeV in $^{16}_\Lambda$O, 
which is consistent with the observed value (see Table \ref{Tab:table1}).
Thus, MBE is represented by MPP+TBA, having only minor effects on the results in this work.

$\Lambda N$ G-matrix interactions $V_{\Lambda N}$ for ESC
are constructed in nuclear matter with Fermi momentum $k_F$~\cite{Yam10}. 
They are represented in coordinate space and parameterized 
in a three-range Gaussian form \cite{Yam10},
\begin{eqnarray}
V_{\Lambda N}(r;k_F)= \sum^3_{i=1}\, (a_i+b_i k_F +c_i k_F^2) \,
\exp {(-r^2/\beta_i^2)} \ .
\label{eq:yng}
\end{eqnarray}
The parameters $(a_i, b_i, c_i)$ are determined so as to simulate the calculated G-matrix 
for each spin-parity state. The procedures to fit the parameters are given in Ref.~\cite{Yam10},
and the determined parameters for ESC are given in Ref.~\cite{PRC89.024310(2014)}. 

Contributions from MBE (MPP+TBA) to G-matrices are represented by modifying 
the second-range parts of $V_{\Lambda N}(k_F,r)$ for ESC by 
$\Delta V_{\Lambda N}(k_F,r)=(a+b k_F+c k_F^2) \exp \left\{ -(r/\beta_2)^2 \right\}$.
It should be noted that the values of parameters $g_P^{(3)}$, $g_P^{(4)}$ and $V_0$
are connected to the values of $a$, $b$ and $c$ through this procedure.
The values of parameters are given in Table~\ref{Gmat-L2}.

\begin{table}
  \caption{Values of parameters in $\Delta V_{\Lambda N}(k_F;r)=  (a+b k_F+c k_F^2) \exp -(r/\beta_2)^2$ with $\beta_2=0.9$ fm.}
  \label{Gmat-L2}
  \begin{ruledtabular}
  \begin{tabular}{ccccc}
   &  $^1E$   & $^3E$    &  $^1O$   &  $^3O$  \\
\hline
 $a$ &    4.809 &    4.345 &    2.701 &    1.611  \\
 $b$ & $-$11.09 & $-$10.57 & $-$7.743 & $-$5.704  \\
 $c$ &    5.264 &    5.035 &    8.004 &    7.599  \\
  \end{tabular}
  \end{ruledtabular}
\end{table}

In applications of nuclear matter G-matrix interactions
$V_{\Lambda N}(r;k_F)$ to finite systems, a basic problem
is how to choose $k_F$ values in each system:
An established manner is to use so called local-density and 
averaged-density approximations $etc$ based on physical insight.
As the better choice to describe $\Lambda$ single particle (s.p.) states, 
we adopt an averaged-density approximation (ADA)~\cite{Yam10}, 
where the averaged value of $k_F$ is defined by 
\begin{eqnarray}
&&k_F =  \left( \frac{3\pi^2 \langle \rho \rangle}{2} \right)^{1/3},
\langle \rho \rangle = \int d^3r \rho_N(\textbf{r}) \rho_\Lambda(\textbf{r}).
\label{ADA}
\end{eqnarray}

In the case of local-density approximation (LDA), $k_F$ values are obtained from
$(\rho_N(\textbf{r})+ \rho_\Lambda(\textbf{r}))/2$ as a function of $\textbf{r}$.
We compare ADA and LDA by calculating $B_\Lambda$ values for $^{89}_{\ \Lambda}$Y
and $^{16}_{\ \Lambda}$O with use of the $\Lambda$-nucleus folding model in which 
$\Lambda N$ G-matrix interactions $V_{\Lambda N}(r;k_F)$ are folded into density 
distributions~\cite{Yam10}. 
For spherical-core systems, the results calculated with the G-matrix folding model 
are similar to those with the HyperAMD used in the following section.
In Table~\ref{Gmat-L3}, the result is shown in the case of using ESC without MBE.
It is demonstrated here that the $B_\Lambda$ values in $^{89}_{\ \Lambda}$Y are reproduced
nicely in both cases of ADA and LDA with no adjustable parameter.
On the other hand, in $^{16}_{\ \Lambda}$O, the value of $B_\Lambda$ obtained with LDA 
is found to be smaller than that with ADA.
Thus, the $B_\Lambda$ values with LDA are similar to (smaller than) those with ADA 
in heavy (light) systems, and eventually the mass dependence of $B_\Lambda$ values 
can be reproduced better using ADA than LDA. 
Hence, the ADA is employed in the present work as an approximate way to use
nuclear matter G-matrix interactions in finite systems.

\begin{table}
\caption{
Values of $B_\Lambda$ in $^{89}_{\ \Lambda}$Y and $^{16}_{\ \Lambda}$O calculated with ADA and LDA (in MeV). 
Observed values of $B_\Lambda$ ($B_\Lambda^{\rm exp}$) are shifted by 0.54 MeV from those in Refs. \cite{PRC64.044302(2001),NPA639.93c(1998)} as explained in Sec. \ref{AMD-BLMD}.
}
  \label{Gmat-L3}
  \begin{ruledtabular}
  \begin{tabular}{cccc}
   & \multicolumn{2}{c}{$-B_\Lambda^{\rm cal}$} & \\
   \cline{2-3}
   &  ADA  & LDA & $-B_\Lambda^{\rm exp}$ \\
\hline
  $^{89}_{\ \Lambda}$Y & $-$23.7  &  $-$23.6 &  $-23.65 \pm 0.10$ \cite{PRC64.044302(2001)} \\
  $^{16}_{\ \Lambda}$O & $-$13.3  &  $-$12.3 &  $-12.96 \pm 0.05$ \cite{NPA639.93c(1998)} \\
  \end{tabular}
  \end{ruledtabular}
\end{table}


\section{Analysis based on HyperAMD}

In this study, we apply the HyperAMD to $p$, $sd$, and $pf$ shell $\Lambda$ hypernuclei, namely from $^{9}_\Lambda$Li up to $^{59}_\Lambda$Fe, in order to describe various structures of these hypernuclei such as an $\alpha$ clustering and prolate, oblate, and triaxial deformations in ground states. Combined with the generator coordinate method (GCM), we perform the systematic analysis of $B_\Lambda$. 

\subsection{Hamiltonian and wave function}

The Hamiltonian used in this study is 
\begin{align}
H = T_{N} + T_{\Lambda} - T_g + V_{NN} + V_{C} + V_{\Lambda N},
\end{align}
where $T_{N}, T_{\Lambda}$, and $T_{g}$ are the kinetic energies of the nucleons, $\Lambda$ particle, and center-of-mass motion, respectively. We use Gogny D1S \cite{Gogny1,Gogny2} as the effective nucleon-nucleon interaction $V_{NN}$, and the Coulomb interaction $V_{C}$ is approximated by the sum of seven Gaussians. As for the $\Lambda N$ interaction $V_{\Lambda N}$, we use the G-matrix interaction discussed above. 

The variational wave function of a single $\Lambda$ hypernucleus is described by the parity-projected wave function, $\Psi^\pm = \hat{P}^\pm \{ \mathcal{A} \{ \varphi_1,\ldots ,\varphi_A \} \otimes \varphi_\Lambda \}$, where
\begin{eqnarray}
\varphi_{i} \propto e^{ - \sum_{\sigma} \nu_\sigma \bigl(r_\sigma - Z_{i\sigma} \bigr)^2 } \otimes (u_i \chi_\uparrow + v_i \chi_\downarrow) \otimes (p \ {\rm or} \ n), \label{varphi}\\
\varphi_\Lambda \propto \sum_{m=1}^M c_m e^{- \sum_{\sigma} \nu_\sigma \bigl(r_\sigma - z_{m\sigma} \bigr)^2} \otimes (a_m \chi_\uparrow + b_m \chi_\downarrow).
\end{eqnarray}
Here the s.p. wave packet of a nucleon $\varphi_{i}$ is described by a single Gaussian, while that of $\Lambda$, $\varphi_\Lambda$, is represented by a superposition of Gaussian wave packets. 
The variational parameters are $\bm{Z}_i$, $\bm{z}_m$, $\nu_\sigma$, $u_i$, $v_i$, $a_m$, $b_m$, and $c_m$.
In the actual calculation, the energy variation is performed under the constraint on the nuclear quadrupole deformation parameters $(\beta, \gamma)$ in the same way as in Ref.~\cite{PRC85.034303(2012)}. By the frictional cooling method, the variational parameters in $\Psi^\pm$ are determined for each set of $(\beta, \gamma)$, and the resulting wave functions are denoted as $\Psi^\pm (\beta,\gamma)$. 

\subsection{Angular momentum projection and generator coordinate method}

After the variation, we project out the eigenstate of the total angular momentum $J$ for each set of $(\beta,\gamma)$ (angular momentum projection; AMP), 
\begin{align}
 \Psi^{J\pm}_{MK}(\beta,\gamma) &=
\frac{2J+1}{8\pi^2}\int d\Omega D^{J*}_{MK}(\Omega)R(\Omega) \Psi^\pm(\beta,\gamma).
\label{eq:AMP}
\end{align}
The integrals over the three Euler angles $\Omega$ are performed numerically. Then the wave functions with differing values of $K$ and $(\beta,\gamma)$ are superposed (generator coordinate method; GCM):
\begin{align}
 \Psi_n^{J\pm}&=\sum_p\sum_{K=-J}^{J} c_{npK} \Psi^{J\pm}_{MK}(\beta_p,\gamma_p). 
 \label{eq:GCM}
\end{align}
The coefficients $c_{npK}$ are determined by solving the Griffin-Hill-Wheeler equation \cite{PRC85.034303(2012)}.
 
\subsection{$B_\Lambda$ and analysis of wave function}
 
The $B_\Lambda$'s are calculated as the energy difference between the ground states of a hypernucleus ($^{A+1}_{\Lambda}Z $) and the core nucleus $(^{A}Z)$ as $B_\Lambda = E(^{A}Z; j^\pm) - E(^{A+1}_{\Lambda}Z; J^\pm)$, where $E(^{A}Z; j^\pm)$ and $ E(^{A+1}_{\Lambda}Z; J^\pm)$ are calculated by GCM. 

We also calculate squared overlap between the $\Psi^{J \pm}_{MK} ( \beta,\gamma)$ and GCM wave function $\Psi^{J \pm}_\alpha$,
\begin{eqnarray}
O^{J\pm}_{MK\alpha} ( \beta,\gamma ) = | \langle \Psi^{J \pm}_{MK} ( \beta,\gamma ) | \Psi^{J \pm}_\alpha \rangle |^2,
\label{Overlap}
\end{eqnarray}
which we call the GCM overlap. 
$O^{J\pm}_{MK\alpha} ( \beta,\gamma )$ shows the contribution of $\Psi^{J \pm}_{MK} ( \beta,\gamma )$ to each state $J^\pm$, which is useful to estimate deformation of each state. In this study, we regard $(\beta, \gamma)$ corresponding to the maximum value of the GCM overlap as nuclear deformation of each state. 

\section{Results and discussions}
  
\begin{table}
  \caption{$-B_\Lambda$ [MeV] calculated with ESC + MBE together with $\langle \rho \rangle$ [fm$^{-3}$] and $k_F$ [fm$^{-1}$] defined by Eq.(\ref{ADA}), and nuclear quadrupole deformation $(\beta, \gamma)$ for each hypernucleus. Values in parentheses are calculated with ESC08c(2012) only in unit of MeV. Observed values $B_\Lambda^{\rm exp}$ are taken from Refs. \cite{NPB52.1(1973),NPA83.306(1979),PRL66.2585(1991),NPA547.369(1992),NPA639.93c(1998),PRC64.044302(2001),NPA754.3(2005),PPNP57.564(2006),PRC90.034320(2014),Gogami}. Values of $B^{\rm exp}_\Lambda$ with dagger are also explained in text.}
  \label{Tab:table1}
  \begin{ruledtabular}
  \begin{tabular}{ccccccc}
 & $\beta$ & $\gamma$ & $\langle \rho \rangle$ & $k_F$ & $-B^{\rm cal}_\Lambda$ & $-B^{\rm exp}_\Lambda$ \\
 \hline
 $^{9}_\Lambda$Li & 0.50 & 2$^\circ$ & 0.072 & 1.02 & $-8.1$($-7.9$) & $-8.50\pm0.12$\cite{NPA754.3(2005)}\\
 $^{9}_\Lambda$Be & 0.87 & 1$^\circ$ & 0.060 & 0.96 & $-8.1$($-7.9$) & $-6.71\pm0.04$\cite{NPB52.1(1973)}\\
 $^{9}_\Lambda$B & 0.45 & 2$^\circ$ & 0.072 & 1.02 & $-8.2$($-8.0$) & $-8.29\pm0.18$\cite{NPA754.3(2005)}\\
 $^{10}_\Lambda$Be & 0.57 & 1$^\circ$ & 0.077 & 1.04 & $-9.0$($-8.7$) & $-9.11\pm0.22$\cite{NPA547.369(1992)},\\
  &  &  &  &  & &  $-8.55\pm0.18$\cite{Gogami}\\
 $^{10}_\Lambda$B & 0.68 & 1$^\circ$ & 0.075 & 1.04 & $-9.2$($-8.9$) & $-8.89\pm0.12$\cite{NPB52.1(1973)}\\
 $^{11}_\Lambda$B & 0.50 & 29$^\circ$ & 0.081 & 1.05 & $-10.1$($-9.8$) & $-10.24\pm0.05$\cite{NPB52.1(1973)}\\
 $^{12}_\Lambda$B & 0.39 & 44$^\circ$ & 0.083 & 1.07 & $-11.3$($-11.0$) & $-11.37\pm0.06$\cite{NPB52.1(1973)},\\
  &  &  &  & &  & $-11.38\pm0.02$\cite{PRC90.034320(2014)}\\
 $^{12}_\Lambda$C & 0.41 & 34$^\circ$ & 0.086 & 1.08 & $-11.0$($-10.7$) & $-10.76\pm0.19$\cite{NPA754.3(2005)}\\
 $^{13}_\Lambda$C & 0.45 & 60$^\circ$ & 0.090 & 1.10 & $-11.6$($-11.3$) & $-11.69\pm0.19$\cite{NPA547.369(1992)}\\
 $^{14}_\Lambda$C & 0.52 & 22$^\circ$ & 0.093 & 1.11 & $-12.5$($-12.4$) & $-12.17\pm0.33$\cite{NPA754.3(2005)}\\
 $^{15}_\Lambda$N & 0.28 & 60$^\circ$ & 0.098 & 1.13 & $-12.9$($-12.6$) & $-13.59\pm0.15$\cite{NPB52.1(1973)}\\
 $^{16}_\Lambda$O & 0.02 &     --     & 0.105 & 1.16 & $-13.0$($-12.7$) & $-12.96\pm0.05$\cite{NPA639.93c(1998)}$^\dag$\\
 $^{19}_\Lambda$O & 0.30 & 3$^\circ$ & 0.110 & 1.18 & $-14.3$($-14.0$) & --\\
 $^{21}_\Lambda$Ne & 0.46 & 0$^\circ$ & 0.106 & 1.16 & $-15.4$($-15.1$) & --\\
 $^{25}_\Lambda$Mg & 0.478 & 21$^\circ$ & 0.116 & 1.20 & $-16.1$($-15.8$) & --\\
 $^{27}_\Lambda$Mg & 0.36 & 36$^\circ$ & 0.125 & 1.23 & $-16.3$($-16.4$) & --\\
 $^{28}_\Lambda$Si & 0.32 & 53$^\circ$ & 0.125 & 1.23 & $-16.6$($-16.4$) & $-17.1\pm0.02$\cite{PPNP57.564(2006)}$^\dag$\\
 $^{32}_\Lambda$S  & 0.23 & 16$^\circ$ & 0.130 & 1.24 & $-17.6$($-17.4$) & $-18.0\pm0.5$\cite{NPA83.306(1979)}$^\dag$\\
 $^{40}_\Lambda$K  & 0.01 &         -- & 0.136 & 1.26 & $-19.4$($-19.2$) & --\\
 $^{40}_\Lambda$Ca & 0.03 &         -- & 0.136 & 1.26 & $-19.4$($-19.2$) & $-19.24\pm1.1$\cite{PRL66.2585(1991)}$^\dag$\\
 $^{41}_\Lambda$Ca & 0.13 & 12$^\circ$ & 0.136 & 1.26 & $-19.6$($-19.4$) & --\\
 $^{48}_\Lambda$K  & 0.01 &         -- & 0.141 & 1.27 & $-20.2$($-20.1$) & --\\
 $^{51}_\Lambda$V  & 0.18 &  2$^\circ$ & 0.151 & 1.31 & $-20.4$($-20.4$) & $-20.51\pm0.13$\cite{PRC64.044302(2001)}$^\dag$\\
 $^{59}_\Lambda$Fe & 0.26 & 23$^\circ$ & 0.142 & 1.28 & $-21.4$($-21.3$) & --\\
  \end{tabular}
  \end{ruledtabular}
\end{table}

\subsection{$B_\Lambda$ in $p$,$sd$, and $pf$ shell $\Lambda$ hypernuclei}
\label{AMD-BLMD}

The calculated values of $B_\Lambda$ for ESC including MBE
are summarized in Table \ref{Tab:table1} together with the values of $k_F$ and $\langle \rho \rangle$, and compared with those calculated by using ESC only (in parentheses) and observed values of $B_\Lambda$ ($B_\Lambda^{\rm exp}$). Here, the $k_F$ values are calculated by Eq.(\ref{ADA}) on the basis of ADA. In Table \ref{Tab:table1}, we also show $(\beta, \gamma)$ which gives the maximum value of the GCM overlap defined by Eq.(\ref{Overlap}).
Recently, in Ref.~\cite{Gogami}, it has been discussed that the $B_\Lambda^{\rm exp}$ measured by the $(\pi^+, K^+)$ experiments are systematically shallower by 0.54 MeV in average than the emulsion data for $^{7}_\Lambda$Li, $^{9}_\Lambda$Be, $^{10}_\Lambda$B and $^{13}_\Lambda$C. It indicates that the reported binding energy of $^{12}_\Lambda$C \cite{NPA754.3(2005)} would be shallower by 0.54 MeV, which is used for the binding energy measurements as the reference in the $(\pi^+, K^+)$ experiments. Therefore, in Table \ref{Tab:table1}, the values of $B^{\rm exp}_\Lambda$ measured by the $(\pi^+, K^+)$ or  $(K^-, \pi^-)$ experiments (with dagger) are shifted by 0.54 MeV deeper from the values reported by Refs.~\cite{NPA83.306(1979),PRL66.2585(1991),NPA639.93c(1998),PRC64.044302(2001),PPNP57.564(2006)}. In spite of this correction, there still remain calibration ambiguities in the $(\pi^+, K^+)$ data. 
One should be careful for this problem, 
when the calculated values of $B_\Lambda$ are compared with these data. 

Let us discuss the calculated values of $B_\Lambda$ shown in Table \ref{Tab:table1}.
As mentioned in Sec. \ref{SecII}, we determine the parameters of MPP and TBA in Eqs. (\ref{eq:2}) and (\ref{eq:4}) so as to reproduce $B_\Lambda^{\rm exp}$ in $^{16}_\Lambda$O in the HyperAMD calculation with ESC + MPP + TBA. 
It is seen that the $B_\Lambda$ with ESC + MPP + TBA reproduces the observed data within about 200 keV except for $^9_\Lambda$Be, $^{15}_\Lambda$N and $^{28}_\Lambda$Si, which is achieved owing to the $k_F$ dependence of the $\Lambda N$ G-matrix interaction used. 
As seen in Table \ref{Tab:table1}, the $k_F$ values become small with decreasing mass number, which means that the $\Lambda N$ G-matrix interaction becomes attractive. The main origin of the $k_F$ dependence is from the $\Lambda N$-$\Sigma N$ coupling terms included in ESC.

\subsection{Effects of core deformation}

For the fine agreement of $B_\Lambda$ values to the experimental data, it is very important to describe properly the core structures, in particular nuclear deformations. 
Recently, many authors have been studying deformations of hypernuclei in $p$-shell \cite{PRC76.034312(2007),PRC78.054311(2008),PRC84.014328(2011),PRC83.044323(2011)}, $sd$-shell \cite{PRC76.034312(2007),PRC78.054311(2008),PTP123.569(2010),PRC84.014328(2011),PRC83.044323(2011),NPA868.12(2011),PRC89.024310(2014),PRC89.044307(2014)}, and $pf$-shell \cite{PRC76.034312(2007),PRC89.024310(2014),PRC89.044307(2014)} mass regions. 
In this study, we take into account deformations of hypernuclei by performing GCM calculations in which intrinsic wave functions with various $(\beta, \gamma)$ deformations $\Psi^\pm (\beta,\gamma)$ are diagonalized. 

\begin{figure}
  \begin{center}
    \includegraphics[keepaspectratio=true,width=86mm]{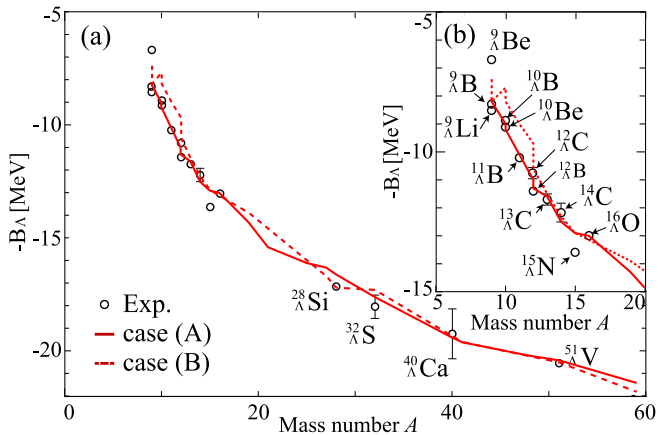}
  \end{center}
  \caption{(Color online) (a) Comparison of $B_\Lambda$ between cases (A) (solid) and (B) (dashed). Open circles show observed values with mass numbers from $A=9$ up to $A=51$, which are taken from Refs. \cite{NPB52.1(1973),NPA83.306(1979),PRL66.2585(1991),NPA547.369(1992),NPA639.93c(1998),PRC64.044302(2001),NPA754.3(2005),PPNP57.564(2006),PRC90.034320(2014)}. $B_\Lambda^{\rm exp}$ measured by $(\pi^+, K^+)$ and $(K^-, \pi^-)$ reactions are shifted by 0.54 MeV as explained in text. (b) Same as (a), but magnified in the $5 \le A \le 20$ region. }
  \label{fig:fig1.eps}
\end{figure}

In order to study the importance of core deformations in the systematic calculations of $B_\Lambda$ values, we perform the GCM calculation by using the spherical wave functions $\Psi^{J\pm}_{MK} (\beta = 0.0)$ in Eq.(\ref{eq:GCM}) (case (B)), whereas Table \ref{Tab:table1} summarizes the GCM results with various deformations (case (A)). 
In the case (B), the $k_F$ value is determined independently from the case (A) with $\Psi^{J\pm}_{MK} (\beta = 0.0)$ by Eq.(\ref{ADA}) for each hypernucleus. 
By using the $k_F$ values determined in the case (B), we also perform the GCM calculations with various $(\beta, \gamma)$ deformations (case (C)). 
Table \ref{Tab:table2} shows the calculated values of $B_\Lambda$ in the cases (A) - (C) in the typical $p$-shell hypernuclei $^{11}_\Lambda$B, $^{12}_\Lambda$B, and $^{13}_\Lambda$C. 
Comparing the cases (A) and (B), we find the considerable discrepancy of $B_\Lambda$, $i.e.$ the $B_\Lambda$ in the case (B) are shallower than those in the case (A), which indicates that the $B_\Lambda$ becomes smaller, if the core nuclei are spherical. 
This is mainly due to the larger $k_F$ value in the case (B) compared with the case (A), 
which comes from the increase of $\langle \rho \rangle$ in a spherical state (see Eq.(\ref{ADA})). 
For example, in the case of $^{12}_\Lambda$B, the obtained value of $B_\Lambda$ is 9.7 MeV with $k_F = 1.16$ fm$^{-1}$ in the case (B), whereas $B_\Lambda = 11.3$ MeV with $k_F = 1.07$ fm$^{-1}$ in the case (A) ($cf.$ $B^{\rm exp}_\Lambda = 11.4 \pm 0.02$ MeV \cite{PRC90.034320(2014)}).
The same difference between the cases (A) and (B) is seen in the other hypernuclei, in particular light hypernuclei with $A < 16$, as shown in Figure \ref{fig:fig1.eps} (a) and (b).

In Table \ref{Tab:table2}, it is also found that the values of $B_\Lambda$ in the case (C) are shallower than those in the case (B), which deviate much from those in the case (A) and the observations. 
This is because the deformation of the core nuclei decreases the overlap between the $\Lambda$ and core nuclei. Since we use the same $k_F$ in the cases (B) and (C), the smaller overlap with deformation in the case (C) makes $B_\Lambda$ shallower. 
Therefore, it can be said that the consistent descriptions of the deformation and the values of $k_F$ determined in deformed states are essential to reproduce the observations. 
$B_\Lambda$ values are given by the balance of two competitive effects: 
(1) The deformation makes the $\Lambda$ s.p. energy ($k_F$ value) shallower (smaller). 
(2) The smaller value of $k_F$ makes the $\Lambda$ s.p. energy deeper due to the density dependence of $\Lambda N$ interaction. In $A >16$ region, generally, deformations make $B_\Lambda$ values smaller because the effect (2) is not so remarkable to cancel the effect (1). On the other hand, in $A <16$ region, deformations make $B_\Lambda$ values larger due to the effect (2). 

\begin{table*}
  \caption{Comparison of $B_\Lambda$ with the cases (A), (B) and (C) in $^{11}_\Lambda$B, $^{12}_\Lambda$B, and $^{13}_\Lambda$C. Value of $k_F$ calculated by Eq.(\ref{ADA}) in each case is also shown. $(\beta, \gamma)$ giving the maximum values of the GCM overlap (Eq.(\ref{Overlap})) are also shown in cases (A) and (C).}
  \label{Tab:table2}
  \begin{ruledtabular}
  \begin{tabular}{cccccccccccc}
              & \multicolumn{3}{c}{$^{11}_\Lambda$B} & \multicolumn{3}{c}{$^{12}_\Lambda$B} & \multicolumn{3}{c}{$^{13}_\Lambda$C} \\
              \cline{2-4} \cline{5-7} \cline{8-10}
              & case (A) & case (B) & case (C) & case (A) & case (B) & case (C) & case (A) & case (B) & case (C)  \\
  \hline
              $-B_\Lambda$ & $-10.1$ & $-9.0$ & $-8.7$ & $-11.3$ & $-9.7$ & $-9.4$ & $-11.6$ & $-11.5$ & $-10.5$  \\
              $k_F$ & 1.05 & 1.13 & 1.13 & 1.07 & 1.16 & 1.16 & 1.10 & 1.15 & 1.15  \\
              $(\beta, \gamma)$ & (0.50,29$^\circ$) &  & (0.50,29$^\circ$) & (0.39,44$^\circ$) &  & (0.39,44$^\circ$) & (0.45,60$^\circ$) &  & (0.45,60$^\circ$) \\
              $-B^{\rm exp}_\Lambda$ & \multicolumn{3}{c}{$-10.24\pm0.05$\cite{NPB52.1(1973)}} 
              & \multicolumn{3}{c}{$-11.37\pm0.06$\cite{NPB52.1(1973)},$-11.38\pm0.02$\cite{PRC90.034320(2014)}} 
              & \multicolumn{3}{c}{$-11.69\pm0.19$\cite{NPA547.369(1992)}} \\
  \end{tabular}
  \end{ruledtabular}
\end{table*}

Let us confirm whether the core deformation is successfully described under the present AMD framework with the Gogny D1S interaction. It can be done essentially by comparing $E2$ transition probabilities $B(E2)$ of the core nuclei with the observations, which are quite sensitive to the nuclear deformation. For example, in $^{12}_\Lambda$B, we calculate $B(E2)$ in $^{11}$B as $B(E2; 5/2^-_1 \to 3/2^-_1) = 16$ $e^2 {\rm fm}^{4}$ by performing the GCM calculation with various $(\beta, \gamma)$ deformations following Refs. \cite{PTP123.303(2010),PRC85.054320(2012)}, which is consistent with the experimental value $B(E2; 5/2^-_1 \to 3/2^-_1) = 14 \pm 3$ $e^2 {\rm fm}^{4}$ \cite{NPA506.1(1990)}. On the basis of the structure calculation for $^{11}$B, we obtain a very reasonable value of $B_\Lambda$ in $^{12}_\Lambda$B by the addition of a $\Lambda$ particle. Then, it is confirmed that our calculations for $B_\Lambda$ are performed in the model space to describe core deformations properly. 

Here, we compare the deformation of hypernuclei with that predicted by Ref.~\cite{PRC78.054311(2008)}, in which $^{13}_\Lambda$C and $^{28}_\Lambda$Si are predicted to be spherical within the framework of relativistic mean-field, whereas the core nuclei $^{12}$C and $^{27}$Si are oblately deformed. It means that the addition of a $\Lambda$ particle makes the core nucleus spherical. 
In the present work, we also find the reduction of the core deformation by the addition of a $\Lambda$ particle. 
However, the degree of deformation change is rather small. 
Thus these hypernuclei are still deformed as shown in Table \ref{Tab:table1}, while $(\beta, \gamma) = (0.50, 59^\circ)$ in $^{12}$C and $(\beta, \gamma) = (0.35, 55^\circ)$ in $^{27}$Si. 
This difference between the present work and Ref.~\cite{PRC78.054311(2008)} mainly comes from the effects by rotational motions, which are included by performing the angular momentum projection (AMP) (see Eq.(\ref{eq:AMP})). 
In fact, it is also found that the deformation of $^{13}_\Lambda$C becomes spherical before performing the AMP \cite{PRC83.044323(2011)}, which is the same trend as predicted by Ref.~\cite{PRC78.054311(2008)}. 
In the present calculation, not only rotational motions but also configuration mixing and shape fluctuations are taken into account by performing the AMP and GCM, which can affect the deformation of hypernuclei.

\subsection{Deviation of $B_\Lambda$ in several hypernuclei}

We comment on the large deviation of $B_\Lambda$ in $^9_\Lambda$Be, $^{15}_\Lambda$N, and $^{28}_\Lambda$Si. 
In $^{9}_\Lambda$Be, it is considered that the Gogny D1S force \cite{Gogny1, Gogny2} overestimates the size of each $\alpha$ particle of 2$\alpha$ cluster structure of the core $^8$Be due to the zero-range density-dependent term, as pointed out by Ref.~\cite{PRC69.034306(2004)}, which would cause the overestimation of $B_\Lambda$ by the decrease of $k_F$ through Eq.(\ref{ADA}). It is found that the $k_F$ value which reproduces the $B_\Lambda^{\rm exp}$ of $^9_\Lambda$Be ($k_F = 1.08$ fm$^{-1}$) is much larger than that shown in Table \ref{Tab:table1} ($k_F = 0.96$ fm$^{-1}$). The smallness of the latter value of $k_F$ is due to the overestimation of the size of $\alpha$ with Gogny D1S. 
It is also found that the same phenomenon appears in the $\Lambda$ hypernuclei with $A < 9$ having $\alpha$ cluster structure by using Gogny D1S. Therefore, we exclude them from being the subject of the present analysis. In such cases, it would be necessary to use appropriate effective $NN$ interactions instead of Gogny D1S. 
In $^{15}_\Lambda$N, the $B_\Lambda^{\rm exp}$ measured by the emulsion experiment \cite{NPB52.1(1973)} seems to be deviating from those of the neighboring hypernuclei in Figure \ref{fig:fig1.eps}(b). This might be due to the difficulties of the analysis and smaller numbers of events in the emulsion experiments. Therefore, we hope to perform a new analysis of the emulsion measurements with large statistic in the future. 
In $^{28}_\Lambda$Si, the value of $B_\Lambda$ is underestimated in case (A), whereas that in case (B) (17.3 MeV) is much closer to the experimental value. 
This might be due to an overestimation of the core deformation, which is seen in the comparison of the electric quadrupole moment $Q$ in the ground state $5/2^+$ of $^{27}$Si, namely, $Q(5/2^+, {\rm AMD}) = 10$ e fm$^2$, whereas $Q(5/2^+, {\rm exp}) = 6.1 \pm 0.4$ e fm$^2$ \cite{STONE2005}. 
Since the calculated values of $k_F$ are almost the same in the cases (A) and (B) (1.23 fm$^{-1}$), the value of $B_\Lambda$ would be in between the values of these cases, if the deformation of $^{27}$Si is smaller than the present result. 

\subsection{$B_\Lambda$ and strength of many-body force}

Finally, we also comment on the relation between $B_\Lambda$ and the strength of MPP and TBA. 
In the present study, the parameters $g^{(3)}_P$ and $g^{(4)}_P$ in Eq.(\ref{eq:2}) 
($V_0$ in Eq.(\ref{eq:4})) are taken as far smaller (less attractive) than those in in Refs. \cite{YFYR13,YFYR14}.
They are determined so as to improve the fitting of $B_\Lambda$ values to the experimental data.
As seen in Table \ref{Tab:table1}, 
the calculated values of $B_\Lambda$ with ESC only reproduce rather well the experimental ones.
Therefore, there remains only a small room to introduce MBE on the basis of ESC. 
On the other hand, in case of MPa \cite{YFYR13,YFYR14}, the parameters of MPP and TBA in hyperonic channels are taken to be the same as those in nucleon channels assuming the stiff EoS of hyperon mixed neutron-star matter. 
It is found that values of $B_\Lambda$ are overestimated if the parameter set of MPa is used combined with ESC. For example, $B_\Lambda$ with MPa are 13.0 MeV for $^{13}_\Lambda$C (\textit{cf.} $B^{\rm exp}_\Lambda = 11.69 \pm 0.19$ MeV), and 14.2 MeV for $^{16}_\Lambda$O (\textit{cf.} $B^{\rm exp}_\Lambda = 12.96 \pm 0.05$ MeV). This indicates that the strength of MPP and TBA in MPa is too strong to reproduce the observations, when MPa is used together with ESC. It is known that two-body $\Lambda N$ effective interactions still have ambiguities, and thus potential depth and $k_F$ dependence are different among models. 
The dependence of MBE on two-body $\Lambda N$ effective interaction models will be discussed in following paper. 
Here, for instance, the strong MPP such as MPa is shown to be allowable in the case of the latest version of ESC08c.

\section{Summary}

On the basis of the baryon-baryon interaction model ESC including MBE, competitive effects of nuclear deformation and density dependence of the $\Lambda N$ interaction are investigated. By using the G-matrix interaction derived from ESC, we perform microscopic calculations of $B_\Lambda$ within the framework of HyperAMD with the ADA treatment for the hypernuclei with $9 \le A \le 59$. 
It is found that the calculated values of $B_\Lambda$ reproduce the experimental data within a few hundred keV,
when the additional density dependence by MBE is taken into account. 
This is achieved by the competition between the nuclear deformation and density dependence of $\Lambda N$ interaction. 
Generally, the overlap between the $\Lambda$ and nucleons varies depending on the degree of core deformation. 
In the light hypernuclei with $A \le 16$, it is found that the $B_\Lambda$ becomes larger by the density dependence of the $\Lambda N$ interaction, because the overlap rapidly decreases for increasing deformation, which mainly comes from the $\Lambda N$-$\Sigma N$ coupling. 
On the other hand, in $sd$-$pf$ shell hypernuclei, the change of the overlap is rather small even if the core deformation is enhanced. Therefore, the density dependence does not affect the $B_\Lambda$ significantly. Instead, increasing deformation makes $B_\Lambda$ smaller by decreasing the overlap. Thus, both of taking into account the core deformations and the treatment of the density dependence of the $\Lambda N$ interaction are essential to understand the systematic behavior of $B_\Lambda$.



\end{document}